\documentclass[showpacs,prb,preprintnumbers,amsmath,amssymb]{revtex4}
\usepackage[latin1]{inputenc}
\usepackage{graphicx}
\usepackage{dcolumn}
\usepackage{bm}
\usepackage[usenames]{xcolor}

\begin{document}

\title{Evidence of a short-range incommensurate $d$-wave charge order from a fermionic two-loop renormalization group calculation of a 2D model with hot spots}
\author{Vanuildo S. de Carvalho$^{1}$}
\author{Hermann Freire$^{1,2}$}
\email{hfreire@mit.edu}
\email[]{hermann\_freire@ufg.br}
\affiliation{$^{1}$Instituto de Física, Universidade Federal de Goiás, 74.001-970, Goiânia-GO, Brazil}
\affiliation{$^{2}$Department of Physics, Massachusetts Institute of Technology, Cambridge, Massachusetts 02139, USA}
\date{\today}

\begin{abstract}
The two-loop renormalization group (RG) calculation is considerably extended here for the two-dimensional (2D) fermionic effective 
field theory model, which includes only
the so-called ``hot spots'' that are connected by the spin-density-wave (SDW) ordering wavevector
on a Fermi surface generated by the 2D $t-t'$ Hubbard model at low hole doping. We compute the
Callan-Symanzik RG equation up to two loops describing the flow of
the single-particle Green's function, the corresponding spectral function, the Fermi velocity, and some of the most important order-parameter
susceptibilities in the model at lower energies.
As a result, we establish that -- in addition to clearly dominant SDW correlations --
an approximate (pseudospin) symmetry relating a short-range \emph{incommensurate} $d$-wave charge order
to the $d$-wave superconducting order indeed emerges at lower energy scales, which is
in agreement with recent works available in the literature addressing the 2D spin-fermion model. 
We derive implications of this possible electronic phase in the ongoing attempt to
describe the phenomenology of the pseudogap regime in underdoped cuprates.
\end{abstract}

\pacs{74.20.Mn, 74.20.-z, 71.10.Hf}

\maketitle

\section{Introduction}

A quantum phase transition \cite{Sachdev5} involving a spin-density-wave (SDW) order taking place in itinerant systems is generally
believed to be crucial to describe a plethora of strongly correlated materials, including, e.g., the cuprates \cite{Taillefer} and the iron-based
pnictide superconductors \cite{Canfield,Goldman}. The canonical approach to describe these transitions is originally due to 
Hertz \cite{Hertz} (later extended by Millis \cite{Millis} and others \cite{Moriya}) and relies upon the assumption that it is possible to integrate out the fermionic excitations
from the microscopic model and to write
down the low-energy theory in terms of an effective action expanded in powers of the order parameter alone.
However, this approach is probably not legitimate for the case of SDW quantum criticality since it has long been argued that it is
potentially dangerous to integrate out
completely the fermions from the microscopic theory in view of the fact that the underlying Fermi surface can experience a dramatic 
reconstruction (see, e.g., Ref. \cite{Sachdev3}) at low energies, thus possibly invalidating the entire approach.

An important step forward consisted in the demonstration by Abanov and Chubukov
that in two spatial dimensions (2D) the Hertz approach for the SDW transition is indeed incomplete and, subsequently,
these authors with collaborators proceeded to formulate the so-called spin-fermion model \cite{Abanov,Abanov2}.
In that work, the high-energy fermions are integrated out in the system such that
they arrive at a low-energy effective theory involving the bosonic SDW order parameter coupled to the fermionic
excitations near the ``hot spots'' (i.e. points in momentum space where the
antiferromagnetic zone boundary intersects the underlying Fermi surface
of the system). This approach has then been extended using the field-theoretical renormalization group (RG)
by Metlitski and Sachdev \cite{Sachdev}
who partly reproduced and partly corrected the results in Refs. \cite{Abanov,Abanov2}. As a result, they
found interesting renormalizations of the Fermi velocity and other physical quantities and confirmed that a breakdown
of Fermi liquid behavior near the hot spots takes place in the system. 

Recently, another important work discussing a slightly modified version of the spin-fermion model from a RG point of view also appeared \cite{Efetov}.
In this work, the authors found, in agreement with some results in Ref. \cite{Sachdev}, that instead 
of a single point separating the antiferromagnet from the
normal metal, there is an intermediate region in the phase diagram that interpolates
between these two phases where the antiferromagnetic long-range order is destroyed but only certain parts of the
Fermi surface are gapped out. This gap is related to the formation of a new quantum state
characterized by a superposition of two nearly degenerate short-range competing orders: a $d$-wave superconducting instability
and a $d$-wave incommensurate charge order. They interpreted this entangled state as possibly 
describing the pseudogap phase at high temperatures observed in underdoped cuprates \cite{Efetov}.
Soon afterwards, Ref. \cite{Sachdev6} has shown that a theory
of these competing orders can describe the very recent X-ray scattering
data obtained in underdoped cuprates \cite{Ghiringhelli,Chang} and in Ref. \cite{Efetov2} a non-linear sigma model theory 
was put forward to describe the phase diagram of these materials in a magnetic field \cite{LeBoeuf}.
In this way, a pressing issue nowadays becomes to understand 
the precise evolution of these fluctuating orders (if possible, in a completely unbiased setting) and also to pinpoint their energy scales in 2D 
quantum critical metals with dominant short-range antiferromagnetic correlations.

For this reason, we will revisit this problem here starting from a slightly different methodological point of view,
but still rooted in a weak-to-moderate coupling perspective: 
instead of integrating out the high-energy fermions to derive an effective theory for the low-energy fermions coupled to the SDW bosonic order parameter field,
we will deal here with only fermionic degrees of freedom in the theory \cite{Shankar,Rice} 
and analyze their effective interactions and, consequently, the potential instabilities in the model 
on equal footing within a two-loop field-theoretical RG framework \cite{Freire,Freire4}. As will become clear, this approach has some technical advantages. Since
it departs directly from the microscopic model, it turns out to be a more unbiased procedure, because the RG method 
itself determines the most important correlations in the model from the corresponding 
flow of the effective couplings and response functions as the RG scale $\Lambda$ is lowered continuously.
We also point out here that this approach applied to the 2D Hubbard model has been shown to reproduce successfully both at 
one-loop \cite{Metzner} and two-loop \cite{Freire4,Katanin} levels an antiferromagnetic phase
near half-filling and the onset of a $d$-wave superconducting phase at larger doping, which agrees qualitatively
with the physics displayed by the cuprate superconductors.

In this paper, we will extend considerably the previous calculation in Ref. \cite{Freire2} and analyze the Callan-Symanzik RG equation up to two loops
for the Green's function of a 2D fermionic model, which includes only the ``hot spots'' that are directly connected by the SDW ordering wavevector
on a Fermi surface generated by the 2D $t-t'$ Hubbard model at low hole doping.
The present fermionic RG approach will allow us to analyze in a direct way the role of several different types of orders 
(that can be either commensurate or incommensurate with the lattice), 
which may or may not emerge in this model at lower energy scales.
As a result, we will show that all the corresponding renormalized parameters flow to an infrared-stable nontrivial fixed point
at two-loop RG level, which 
controls the universal physics of the model at large time scales and long distances. We will explore here the implications
of such a nontrivial fixed point in the model. As will become clear shortly, this fixed point implies that: 

(a) Non-Fermi liquid behavior is obtained
near the hot spots, displaying an emergent approximate (pseudospin) symmetry
relating a short-range incommensurate $d$-wave charge order to the $d$-wave superconducting instability.

(b) The single-particle renormalized Green's function, the resulting spectral function, and the tunneling density of states
of the model should all obey precise scaling forms
which can be ultimately verified experimentally using, for instance, angle-resolved photoemission spectroscopy (ARPES) 
and scanning tunneling microscope (STM) probes. 

Finally, we put all of our present results into context with other very recent data
obtained in the literature addressing the 2D spin-fermion model.

\begin{figure}[t]
 \includegraphics[height=2.4in]{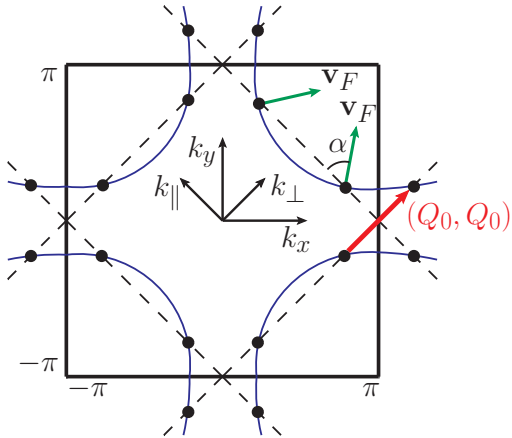}
 \caption{(Color online) The 2D fermionic model consisting of eight ``hot spots'' on the Fermi surface
 which are directly connected by the commensurate spin density wave (SDW) ordering wavevector $\mathbf{Q}=(\pi,\pi)$. We also show the Fermi velocities at some
 of these points together with the angle $\alpha$. The incommensurate vector $\mathbf{\widetilde{Q}}=(Q_0,Q_0)$ is also displayed.}
 \label{fig:Fermi_surface}
\end{figure}

\section{The Model}

Our starting point is the noninteracting part of the model with the energy dispersion given
by $\xi_{\mathbf{k}}=-2t(\cos(k_x)+\cos(k_y))-4t'\cos(k_x)\cos(k_y)-\mu$
with $t$ and $t'$ being, respectively, the nearest neighbor and next-nearest neighbor hoppings and $\mu$ is the
chemical potential. For the cuprates, the appropriate choice of parameters is $t'=-0.3t$, which, at low hole doping, results in the curved Fermi
surface (FS) shown in Fig. 1. This FS intersects the antiferromagnetic zone boundary at eight points (i.e., the ``hot spots'').
The eight ``hot spots'' are displayed in Fig. \ref{fig:Fermi_surface}. If we rotate the momentum axes $(k_x,k_y)$ by $45^{\circ}$,
we can define the new axes $(k_{\parallel},k_{\perp})$, where the momenta $k_{\parallel}$ and $k_{\perp}$ refer, respectively, to the momentum parallel
and normal to half of the hot spots on the FS (we mention here that,
for the other half of the hot spots, the roles of $k_{\parallel}$ and $k_{\perp}$
are simply interchanged). Moreover, since we
will be interested in the universal properties of this
model, we shall linearize the energy dispersion around the FS as $\xi_{\mathbf{k}}\approx v_{\perp} (|k_{\perp}|-k_{F}^{\perp})-
v_{\parallel} (|k_{\parallel}|-k_{F}^{\parallel})+ \mathcal{O}[(\mathbf{k}-\mathbf{k}_F)^{2}]$
with the normal and parallel components of the Fermi velocity $\mathbf{v_F}$=($v_{\parallel}$,$v_{\perp}$) given, respectively, by 
$v_{\perp}=|\nabla_{(k_{\parallel},k_{\perp})}\xi_{\mathbf{k}}|_{\mathbf{k}=\mathbf{k}_F}|
\sin{\alpha}$ and $v_{\parallel}=|\nabla_{(k_{\parallel},k_{\perp})}\xi_{\mathbf{k}}|_{\mathbf{k}=\mathbf{k}_F}|
\cos{\alpha}$, where 
$\mathbf{k}_F$ is the Fermi momentum at the hot spots and $\alpha\approx 64.4^{\circ}$ is the angle of the Fermi velocity on 
the rotated momentum axes for the cuprates at $4\%$ of hole doping (i.e. $\mu\approx -0.77t)$. Therefore,
initially, one obtains that $v_{\perp}\approx 1.7t$, $v_{\parallel}\approx 0.8t$ and $(v_{\perp}/v_{\parallel})\approx 2.1$.
Both the momenta parallel to the FS and perpendicular to the FS are restricted to the interval $[-k_{c},k_{c}]$, where $k_c$
essentially determines the ultraviolet (UV) momentum cutoff in our theory. This implies also an energy cutoff which is given by $\Lambda_0=2v_{F}k_{c}$ which
we choose to be equal to the full bandwidth of the problem, i.e. $\Lambda_0=8t$.

\begin{figure}[t]
 \includegraphics[height=2.7in]{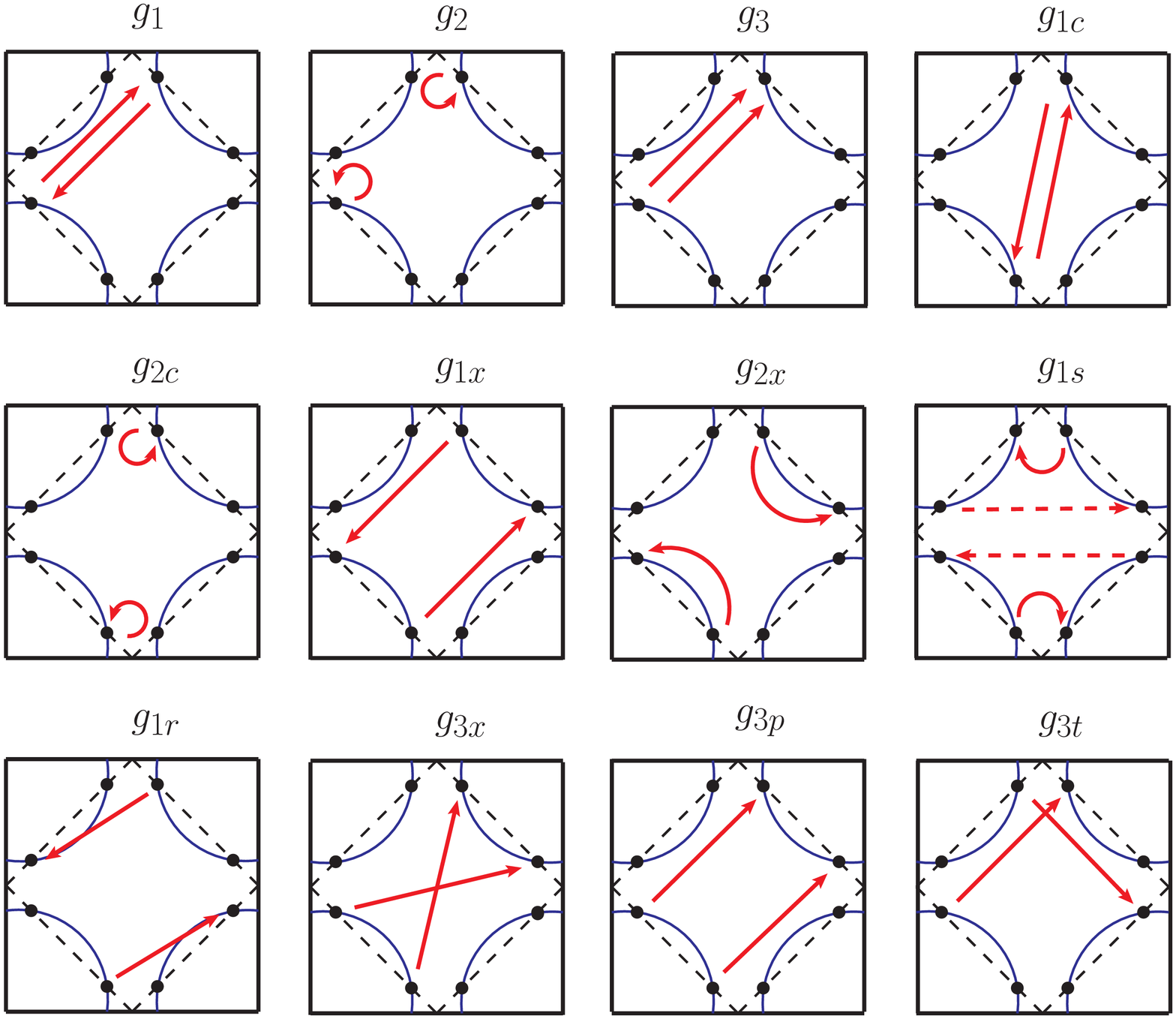}\\
 \includegraphics[height=0.85in]{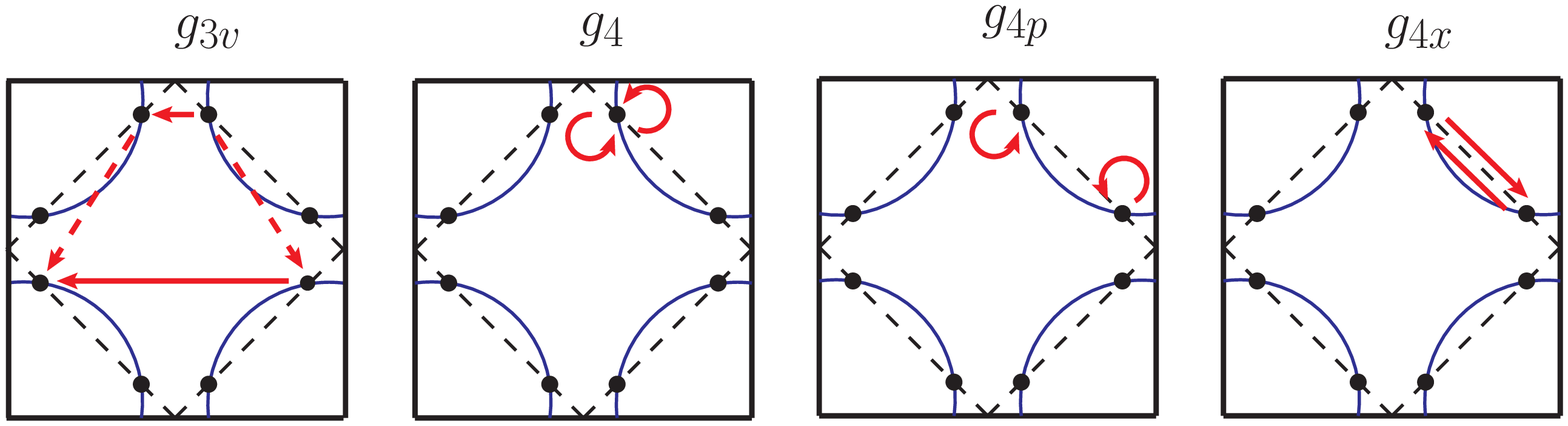}
 \caption{(Color online) Some relevant couplings in the present model. We follow a ``g-ology'' notation adapted to our 2D problem at hand.
For the $g_{1s}$ and $g_{3v}$ couplings, we show two possible scattering
processes (one by a solid line and the other by a dashed line)
which are in fact always equal in our two-loop RG results.}
 \label{fig:Interactions}
\end{figure}

The model at $T = 0$ and constant
chemical potential $\mu = E_F$ becomes described by the 
partition function $\mathcal{Z}=\int \mathcal{D}[\overline{\psi},\psi] \exp({i\int_{-\infty}^{\infty} dt\, L_{R}[\overline{\psi},\psi]})$ with the
fully interacting renormalized Lagrangian $L_{R}$ given by

\vspace{-0.2cm}

\begin{align}\label{lagrangian}
L_{R}&=\sum_{\mathbf{k},\sigma}Z\,\overline{\psi}_{R\sigma}(\mathbf{k})\big[i\partial_t-Z_{v_{\perp}}Z^{-1}v_{\perp R}(|k_{\perp}|-k^{\perp}_{F})\nonumber\\
&+Z_{v_{\parallel}}Z^{-1}v_{\parallel R}(|k_{\parallel}|-k^{\parallel}_{F})\big]\psi_{R\sigma}(\mathbf{k})\nonumber\\
&-\sum_{i}\sum_{\substack{{\mathbf{k_1,k_2,k_3}} \\ {\sigma,\sigma'}}}
Z^{2}g_{i,B}\,\overline{\psi}_{R\sigma}(\mathbf{k_4})\overline{\psi}_{R\sigma'}(\mathbf{k_3})\psi_{R\sigma'}(\mathbf{k_2})\psi_{R\sigma}(\mathbf{k_1})
\end{align}

\noindent where $\mathbf{k_4}=\mathbf{k_1}+\mathbf{k_2}-\mathbf{k_3}$ and the volume $V$ has been set equal to unity. 
The bare quantities (denoted by the index $B$) are related to the renormalized quantities (denoted by the index $R$) by the following expressions: 
$\psi_{B\sigma}(\mathbf{k})= Z^{1/2}\psi_{R\sigma}(\mathbf{k})$, 
$\overline{\psi}_{B\sigma}(\mathbf{k})= Z^{1/2}\overline{\psi}_{R\sigma}(\mathbf{k})$, 
$v_{\perp B}=Z_{v_{\perp}}Z^{-1}v_{\perp R}$ and $v_{\parallel B}=
Z_{v_{\parallel}}Z^{-1}v_{\parallel R}$, where $Z$ is the 
quasiparticle weight. The renormalized Grassmann fields
$\overline{\psi}_{R\sigma}(\mathbf{k})$ and $\psi_{R\sigma}(\mathbf{k})$ are associated, respectively, to the creation and annihilation operators
of excitations lying in the vicinity of the hot spots with momentum $\mathbf{k}$ and spin projection $\sigma$. The index $i$ runs over possible 
interaction processes in
the model that produce logarithmic divergences within perturbation theory, i.e., 
$i=1,2,3,1c,2c,1x,2x,1s,1r,3x,3p,3t,3v,4,4p,4x$ (for details
of the couplings taken into account, see Fig. 2). In this way, to keep a close connection with other RG works in the literature,
we follow a ``g-ology'' notation \cite{Solyom}, adapted to our 2D problem at hand.
Moreover, we must define the dimensionless renormalized couplings of the model -- which we will denote simply by $g_{iR}$ --
in the following way: $g_{i,B}=N_{0}^{-1}Z^{-2} \left[{g}_{iR}+\delta {g}_{iR}\right]$, where $D(0)=N_0/2$
is the density of states at the Fermi level. (We point out here that 
the renormalized dimensionless couplings $g_{iR}$ should not be confused with the bare coupling constants $g_{i,B}$ of the model, which are in turn dimensionful.)
In all the above expressions, we set 
conventionally $Z=1+\delta Z$, $Z_{v_{\perp}}=1+\delta Z_{v_{\perp}}$, 
$Z_{v_{\parallel}}=1+\delta Z_{v_{\parallel}}$, where $\delta Z$, $\delta Z_{v_{\perp}}$,
$\delta Z_{v_{\parallel}}$ and $\delta {g}_{iR}$ are
the so-called counterterms that must be determined consistently within the  
renormalized perturbation theory \cite{Peskin} (see next section).
For simplicity and to keep the total number of Feynman diagrams to be computed in this work not extremely large, 
we shall neglect from this point on the interaction processes described by the couplings $g_{4}$, $g_{4p}$ and $g_{4x}$. As will become
clear shortly, those specific scattering processes will only generate logarithmic divergent diagrams at 
two-loop order or beyond. Because of this crucial fact, experience
with one-dimensional systems\cite{Solyom} indicates that, within a perturbative regime, those interactions 
are not expected to alter qualitatively our present results.

\section{RG Strategy}

The methodology of our RG scheme follows closely the standard field-theoretical approach\cite{Peskin,Weinberg2}, which 
was also explained in full detail in several of the authors' previous papers with collaborators \cite{Freire,Freire2,Freire3}. 
If one applies a naive perturbation theory for the present model, 
divergences (or non-analyticities) emerge at lower energies 
at the calculation of several
important quantities of the model such as vertex corrections, self-energy, and linear response functions. This result 
normally implies that the bare perturbation theory setup is not appropriate for this case, since it is known to be written
in terms of the bare parameters (defined at the microscopic scale $\Lambda_0$) 
and not the low-energy quantities of the model. As was shown before, we circumvent this problem
by rewriting all the bare parameters of the theory in terms of the corresponding renormalized ones plus additional counterterms. 
The main role of these counterterms is to regulate 
the theory at a floating RG scale $\Lambda$ and, in general, they must be calculated order by order in perturbation theory. 
By doing this, the newly-constructed renormalized perturbation theory becomes a well-defined expansion in terms of the 
renormalized parameters and, in this way, its predictions can be compared to experiments. Since this program is successfully accomplished here, the 
model is renormalizable.

We can divide the RG flow obtained in this work into two different energy regimes: 
if the RG scale $\Lambda$ is such that $\Lambda>E_{\mu}$ (i.e. high energies) --
where $E_{\mu}$ is an energy scale that in our present problem 
turns out to be a bit larger than the modulus of the chemical potential $|\mu|$ -- 
one can assume that the hot spots in the model exhibit approximate 
nesting at the commensurate SDW ordering wavevector $\mathbf{Q}=(\pi,\pi)$ in the sense that 
the particle-hole bubble for this case becomes almost logarithmic divergent as a function of $\Lambda$.
The reason we include this channel in our present theory comes from the experimental observation that in underdoped cuprates
there is a transition from a metallic
paramagnet to a metallic antiferromagnet, despite the initial absence of nesting at $\mathbf{Q}=(\pi,\pi)$ of the underlying Fermi surface of these materials.
A similar thing happens for the more complicated two-loop contributions in our theory (to be discussed in a more detailed way later in this paper), 
which also become nearly logarithmic divergent for this high-energy regime.
By contrast, the hot spots
at the incommensurate wavevector $\mathbf{\widetilde{Q}}=(Q_0,Q_0)$ (see Fig. \ref{fig:Fermi_surface}) 
exhibit initially good nesting, which implies that the particle-hole channel at this
wavevector is already logarithmic divergent as a function of the RG scale $\Lambda$. Lastly, we point out that  
the particle-particle channel at $\mathbf{q}=(0,0)$ in the model 
is also logarithmic divergent, since the Fermi surface exhibits of course parity symmetry.

If we consider, however, the other regime $\Lambda\lesssim E_{\mu}$ (i.e. low energies), 
then it becomes clear from the previous discussion that one cannot assume any longer that the 
hot spots exhibit approximate nesting 
at $\mathbf{Q}=(\pi,\pi)$ and the two-loop contributions to the RG flow also become irrelevant. Indeed, in this case only the remaining 
channels (i.e. the particle-hole bubble at $\mathbf{\widetilde{Q}}=(Q_0,Q_0)$ and the particle-particle bubble at $\mathbf{q}=(0,0)$) 
turn out to be logarithmic divergent as function of $\Lambda$. 
We will concentrate throughout this work
on the first regime though (i.e. $\Lambda>E_{\mu}$), which we believe is more relevant as a weak-coupling signature of the intermediate to strong
coupling high-temperature pseudogap phase that is widely observed in the hole-doped cuprates. In this case, 
all particle-hole and particle-particle one-loop channels discussed above
together with the more complicated two-loop contributions compete with one another in the RG flow. 
Therefore, for this high-energy regime, the two-loop fermionic RG method turns out to be an ideal tool to describe 
the universal properties of such system
from a weak coupling perspective in view of its already-advertised unbiased nature.

\subsection{One-loop RG}

In Fermi liquid theory, interactions between the fermions can in general give rise to either density waves (DW)
or superconducting (SC) instabilities at low energies. Those instabilities are related to divergences
in the corresponding response functions and may take place when these enhancements eventually become larger than the bare
interactions in the appropriate channel. As was explained before, in setting up a conventional perturbation theory at $T=0$ to perform
calculations with the present model, one encounters for high energies (i.e. $\Lambda>E_{\mu}$) logarithmic divergent 
particle-hole ($\Pi_{ph}$) and particle-particle ($\Pi_{pp}$) bubbles at one-loop order, which are given, respectively, by

\vspace{-0.2cm}

\begin{align}
\Pi_{ph}(q_0=\Lambda,\mathbf{Q})&=\int_{\mathbf{k},\omega}G_{0}(\mathbf{k},\omega)G_{0}(\mathbf{k+Q},q_0+\omega)\nonumber\\
&\approx i \frac{N_0}{2} \ln\left(\frac{\Lambda_0}{\text{max}\{\Lambda,E_{\mu}\}}\right),
\end{align}

\vspace{-0.2cm}

\begin{align}
\Pi_{ph}(q_0=\Lambda,\mathbf{\widetilde{Q}})&=\int_{\mathbf{k},\omega}G_{0}(\mathbf{k},\omega)G_{0}(\mathbf{k+\widetilde{Q}},q_0+\omega)\nonumber\\
&= i \frac{N_0}{2} \ln\left(\frac{\Lambda_0}{\Lambda}\right),
\end{align}

\vspace{-0.2cm}

\begin{align}
\Pi_{pp}(q_0=\Lambda,\mathbf{q}=0)&=\int_{\mathbf{k},\omega}G_{0}(\mathbf{k},\omega)G_{0}(\mathbf{-k},-\omega+q_0)\nonumber\\
&= -i \frac{N_0}{2} \ln\left(\frac{\Lambda_0}{\Lambda}\right),
\end{align}

\noindent where $\int_{\mathbf{k},\omega}=\int\frac{d^{2}\mathbf{k}}{(2\pi)^{2}}\frac{d\omega}{2\pi}$ and $G_{0}(\mathbf{k},\omega)$ is
the noninteracting Green's function. An important point we wish to stress here is that, 
despite the fact that all results presented in this work are done at $T=0$, 
the RG scale $\Lambda$ can also be interpreted, to logarithmic accuracy, as playing the role of a finite temperature
$T$ in the system. In other words, the high-energy regime ($\Lambda>E_{\mu}$) would correspond physically to high temperatures.

\begin{figure}[t]
 \includegraphics[height=2.4in]{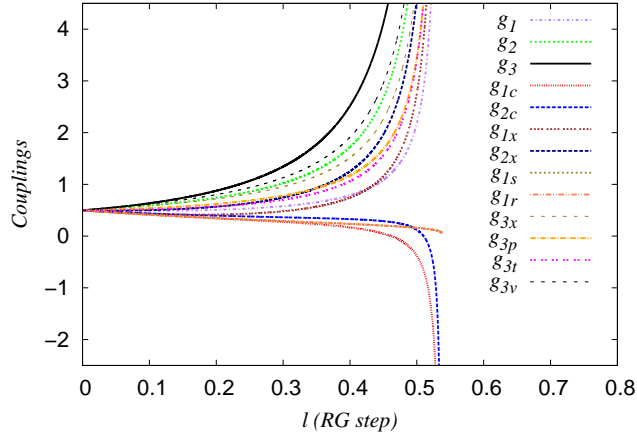}
 \caption{(Color online) One-loop RG flow of all renormalized dimensionless couplings in the present model for the initial conditions $g_{iR}{(0)}=0.5$.}
 \label{fig:Couplings_One_Loop}
\end{figure}

The Feynman diagrams and the one-loop RG equations can be easily found in Fig. \ref{fig:Vertex_corrections} and in the 
RG equations shown in Appendix A (in this last case, by keeping only the terms in the beta function which are quadratic in the couplings). 
Then, we proceed to solve these equations numerically by using the standard
fourth-order Runge-Kutta method. The resulting one-loop RG flow for the dimensionless couplings is shown in Fig. \ref{fig:Couplings_One_Loop}
for the initial condition $g_{iR}{(0)}=0.5$ (for $i=1,2,3,1c,2c,1x,2x,1s,1r,3x,3p,3t,3v$). 
This would correspond roughly to a local on-site Hubbard interaction $U$ given by $(U/\Lambda_0)\approx 0.45$ (see next section).
The numerical solution of these one-loop RG equations shows that, even though there is no evidence of a non-trivial fixed point
in this model at this order, almost all dimensionless coupling constants $g_{iR}$ 
diverge at the same critical RG step value $l_c=\ln(\Lambda_0/\Lambda_c)$, which in turn depends only on the initial conditions for the couplings. 
This divergence is appropriately captured by the following scaling ansatz

\vspace{-0.3cm}

\begin{equation}\label{ansatz}
 g_{iR}(l)=\frac{C_{i}}{l_{c}-l},
\end{equation}

\noindent where the RG step is $l=\ln(\Lambda_0/\Lambda)$ and the $C_{i}$ are universal constants that do not depend on the initial conditions. 
Substituting Eq. \eqref{ansatz} into the one-loop RG equations, we obtain a set of
thirteen polynomial equations for the $C_{i}$, which could be 
written in short form as  $C_{i}+\beta_{i}^{1-loop}(\{C_{j}\})=0$, where $\beta_{i}^{1-loop}=\Lambda d g_{i R}/d\Lambda$. The solution of this set of equations is 
obtained numerically and corresponds to 

\vspace{-0.3cm}

\begin{equation}\label{fixed_point_one_loop}
 \begin{bmatrix} C_{1} \\ C_{2} \\ C_{3} \\ C_{1c} \\ C_{2c} \\ C_{1x} \\ C_{2x} \\ C_{1s} \\ C_{1r} 
 \\ C_{3x} \\ C_{3p} \\ C_{3t} \\ C_{3v} \end{bmatrix}\approx \begin{bmatrix} 0.073 \\ 0.234 \\ 0.393 
 \\ -0.039 \\ -0.020 \\ 0.127 \\ 0.186 \\ 0.0 \\ 0.0 \\ 0.185 \\ 0.108 \\ 0.124 \\ 0.249 \end{bmatrix}.
\end{equation}

\noindent We observe from the one-loop RG result that the leading coupling constants are represented by Umklapp interactions.
As will become clear shortly, those interactions will favor SDW antiferromagnetic ordering tendencies, 
which will be the dominant correlations in the system.
By contrast, upon inclusion of quantum fluctuation effects, the Cooper-pair interaction processes $g_{1c}$ and $g_{2c}$ -- that are initially taken to be 
repulsive -- flow naturally to attractive values. These latter
couplings will enhance, e.g., $d$-wave pairing correlations which will of course tend to emerge as a possible competing order in the system.

\subsection{Two-loop RG}

As we emphasized in the previous sections, we will concentrate throughout this work only in the physical regime, 
in which the energies are actually larger than a typical scale $E_{\mu}$
(i.e. $\text{max}\{|p_0|,
v_{\perp R}(|p_{\perp}|-k^{\perp}_{F}),v_{\parallel R}
(|p_{\parallel}|-k^{\parallel}_{F})\}> E_{\mu}$) associated with an initial absence of 
perfect nesting at $\mathbf{Q}=(\pi,\pi)$ displayed by the underlying Fermi surface of the model.
In this high-energy regime, if we calculate the renormalized self-energy of the model at $T=0$ up to two-loop order for $v_{\perp R}\gg v_{\parallel R}$, we 
obtain that its non-analytic contribution (say, with 
$p_{\parallel}>0$ and $p_{\perp}>0$) is given approximately by

\vspace{-0.3cm}

\begin{align}\label{self-energy}
&-i\Sigma_{R}(p_0,\mathbf{p})\approx \frac{i\gamma}{8}\big[p_{0}-v_{\perp R}(p_{\perp}-k^{\perp}_{F})+v_{\parallel R}
(p_{\parallel}-k^{\parallel}_{F})\big]\nonumber\\
&\times\biggl\{\ln\biggl[\frac{p_{0}-2v_{\perp R}k_{c}-v_{\perp R}(p_{\perp}-k^{\perp}_{F})+v_{\parallel R}(p_{\parallel}-k^{\parallel}_{F})+i\delta}{p_{0}
-v_{\perp R}(p_{\perp}-k^{\perp}_{F})+v_{\parallel R}(p_{\parallel}-k^{\parallel}_{F})+i\delta}\biggr]\nonumber\\
&+\ln\biggl[\frac{p_{0}+2v_{\perp R}k_{c}-v_{\perp R}(p_{\perp}-k^{\perp}_{F})+v_{\parallel R}(p_{\parallel}-k^{\parallel}_{F})-i\delta}{p_{0}
-v_{\perp R}(p_{\perp}-k^{\perp}_{F})+v_{\parallel R}(p_{\parallel}-k^{\parallel}_{F})-i\delta}\biggr]\biggr\}\nonumber\\
&+i\big[\delta Z\,p_{0}-\delta Z_{v_{\perp}}v_{\perp R}(p_{\perp}-k^{\perp}_{F})+\delta Z_{v_{\parallel}}v_{\parallel R}
(p_{\parallel}-k^{\parallel}_{F})\big],
\end{align}

\noindent where $\delta\rightarrow 0^{+}$ and $\gamma=(g^{2}_{1R}+g^{2}_{2R}+g^{2}_{1cR}+g^{2}_{2cR}+g^{2}_{1xR}+g^{2}_{2xR}
-g_{1R}g_{2R}-g_{1cR}g_{2cR}-g_{1xR}g_{2xR}-g_{3pR}g_{3xR}+g^{2}_{3pR}
+g^{2}_{3xR}+\frac{g^{2}_{3R}}{2})$. In addition, the density of states at the Fermi level in the above case  
could be well approximated by $D(0)\approx k_c/(2\pi^2 v_{\perp R})$. 
Using a standard RG condition (see Appendix B for the details), we obtain that
the anomalous dimension defined by $\eta=\Lambda \,d\ln Z/d\Lambda$ is given by $\eta=\gamma/4$.
In an analogous way, we obtain that the quantities defined by $\eta_{v_{\perp}}=\Lambda\, d\ln Z_{v_{\perp}}/d\Lambda$ and 
$\eta_{v_{\parallel}}=\Lambda\, d\ln Z_{v_{\parallel}}/d\Lambda$ are given by
$\eta_{v_{\perp}}=\eta_{v_{\parallel}}=\gamma/4$. This latter result implies that
the Fermi velocity is not renormalized up to two-loop order in the
present model within the high-energy regime. This should be contrasted with
the 2D spin-fermion model where a different renormalization for the Fermi velocity is obtained \cite{Abanov2,Sachdev}.

\begin{figure}[t]
 \includegraphics[height=2.4in]{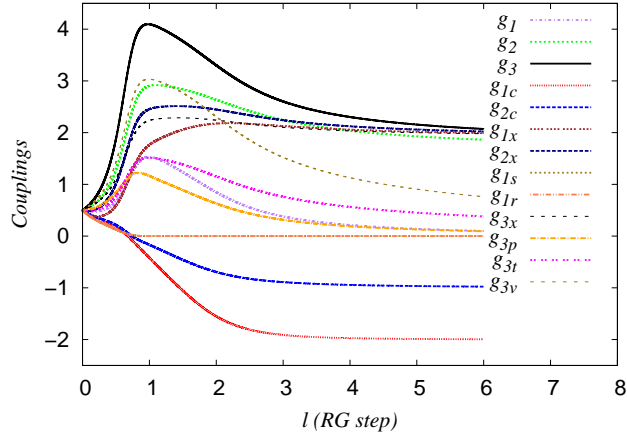}
 \caption{(Color online) Two-loop RG flow of the renormalized dimensionless couplings in the present model for the initial conditions $g_{iR}{(0)}=0.5$.}
 \label{fig:Couplings_two_Loops}
\end{figure}

Following once again the field-theoretical RG methodology, we obtain the two-loop flow equations
for the renormalized dimensionless couplings $g_{iR}$, which are fully shown in Appendix A. We solve those RG equations numerically using
the same numerical procedure explained before. The resulting
two-loop RG flow for the dimensionless couplings is shown in Fig. \ref{fig:Couplings_two_Loops}
using the same initial condition as in the previous section $g_{iR}{(0)}=0.5$ (for $i=1,2,3,1c,2c,1x,2x,1s,1r,3x,3p,3t,3v$),
which corresponds to a local on-site Hubbard interaction $U$ given by $(U/\Lambda_0)\approx 0.45$. 
We confirm here that the inclusion of two-loop order quantum fluctuations 
completely removes the divergence at a finite critical energy scale $\Lambda_c$, which was obtained in one-loop RG order. 
As a result, the renormalized couplings now approach asymptotically a nontrivial infrared fixed point,
which is given by

\vspace{-0.3cm}

\begin{equation}\label{fixed_point_two_loop}
\begin{bmatrix} g^{*}_{1} \\ g^{*}_{2} \\ g^{*}_{3} \\ g^{*}_{1c} \\ g^{*}_{2c} 
\\ g^{*}_{1x} \\ g^{*}_{2x} \\ g^{*}_{1s} \\ g^{*}_{1r} \\ g^{*}_{3x} \\ g^{*}_{3p} 
\\ g^{*}_{3t} \\ g^{*}_{3v} \end{bmatrix}\approx \begin{bmatrix} 0.0 \\ 1.684 \\ 1.841 
\\ -2.0 \\ -1.0 \\ 1.918 \\ 1.918 \\ 0.0 \\ 0.0 \\ 1.918 \\ 0.0 \\ 0.0 \\ 0.0 \end{bmatrix}.
\end{equation}

\noindent We mention here that the above result does not depend
on our numerical choice for the initial coupling constants (as long as the nontrivial fixed point is reached within the high-energy regime)
and, for this reason, it implies a new universality class in the present problem.
It is also interesting to note that many couplings flow asymptotically to zero,
whereas only some flow to nonzero values. This fact could potentially simplify
the solution of this model by means of other methods. 

In order to study the stability of this nontrivial 
fixed point in the coupling parameter space, one must analyze the RG flow in
the vicinity of this point \cite{Weinberg2}. We can thus write the two-loop RG flow equations as 
\begin{equation}\label{approximation_two_loop}
\Lambda\frac{d}{d\Lambda}[g_{iR}(\Lambda)-g_{i}^{*}]=\sum_{j}M_{ij}[g_{jR}(\Lambda)-g_{j}^*],
\end{equation}

\noindent where $M_{ij}=\frac{\partial\beta_{i}(\{g_{jR}\})}{\partial g_{jR}}\vert_{g_{R}=g^{*}}$ is a matrix 
defined at the nontrivial fixed point $g^{*}=(g^{*}_{1},g^{*}_{2},g^{*}_{3},\ldots,g^{*}_{3v})$, and 
$\beta_{i}(\{g_{jR}\})$ correspond to the RG beta functions up to two-loop order. The solution of 
Eq. \eqref{approximation_two_loop} can be expanded in eigenvectors of this matrix as 

\vspace{-0.3cm}

\begin{equation}\label{solution_two_loop}
g_{iR}(\Lambda)=g_{i}^{*}+\sum_{j}b_m V_{j}^{i}\Lambda^{m_j},
\end{equation}

\noindent where $V_{j}$ is an eigenvector of $M_{ij}$ with associated eigenvalue $m_j$ 
and $b_m$ are a set of expansion coefficients. Diagonalizing the matrix $M_{ij}$, 
we determine numerically that its eigenvalues are indeed all positive and given by

\vspace{-0.4cm}

\begin{equation}\label{eigenvalues_two_loop}
 \begin{bmatrix} m_{1} \\ m_{2} \\ m_{3} \\ m_{1c} \\ m_{2c} 
\\ m_{1x} \\ m_{2x} \\ m_{1s} \\ m_{1r} \\ m_{3x} \\ m_{3p} 
\\ m_{3t} \\ m_{3v} \end{bmatrix} \approx \begin{bmatrix} 15.72 \\ 8.05 \\ 7.91 \\ 7.60 \\ 7.30 \\ 5.91 \\ 5.90 \\ 5.78 \\ 5.67 \\ 5.67 \\ 3.59 \\ 2.76 \\ 0.08 \end{bmatrix}.
\end{equation}

\noindent This result establishes that any trajectory in the coupling parameter space 
close to the fixed point $g^{*}$ converges to it as the energy scale $\Lambda$ is lowered. 
Therefore, $g^{*}$ corresponds to an infrared-stable nontrivial fixed point with its basin of attraction 
having the same dimension of the coupling space of the 2D fermionic model. As a result,
this fixed point controls the universal physics of the model at large time scales and long distances. 
In the next section, we will begin to explore the implications
of such a nontrivial fixed point in the model.

\section{Callan-Symanzik Equation}

We can use the two-loop RG also to calculate the renormalized Green's function close to the nontrivial fixed point obtained above. 
Since $G_{R}(p_0,\mathbf{p},\{g_{R}\};\Lambda)=
Z^{-1}(\{g_{R}\};\Lambda)G_{B}(p_0,\mathbf{p},\{g_{B}\})$, one can see that $G_{R}$ satisfies the Callan-Symanzik equation \cite{Callan,Symanzik,Peskin}

\vspace{-0.4cm}

\begin{align}\label{CZE}
&\left(\Lambda\frac{\partial}{\partial \Lambda}+\sum_{i}\beta_{i}(\{g_{i R}\})\frac{\partial}{\partial g_{i}}+\eta\right) G_{R}(p_0,\mathbf{p},\{g_{R}\};\Lambda)=0,
\end{align}

\noindent where the two-loop RG beta functions $\beta_{i}=\Lambda d g_{i R}/d\Lambda$ are fully given in Appendix A. 
On dimensional grounds, $ G_{R}(p_0,\mathbf{p},\{g_{R}\};\Lambda)$ must
also satisfy the following equation \cite{Weinberg,Ferraz}: $\left(\Lambda\partial/\partial\Lambda+p_{0}\partial/\partial p_{0}\right) 
G_{R}(p_0,\mathbf{p=k_F};\Lambda)=-G_{R}(p_0,\mathbf{p=k_F};\Lambda)$ which, at the fixed point, implies that

\vspace{-0.4cm}

\begin{align}\label{Grenorm}
&\left(p_{0}\frac{\partial}{\partial p_{0}}+1-\eta^{*}\right) G_{R}(p_0,\mathbf{p=k_F};\Lambda)=0.
\end{align}

\noindent After performing analytic continuation, the above equation has the solution 

\vspace{-0.4cm}

\begin{align}
G_{R}(p_0+i0^{+},\mathbf{p=k_F};\Lambda)=\frac{1}{\Lambda}\left(\frac{\Lambda}{p_0+i0^{+}}\right)^{1-\eta^{*}}. 
\end{align}

\noindent Since the the analytical structure of the renormalized Green's function of the model exhibits no quasiparticle form,
this implies a complete breakdown of Fermi liquid behavior 
near the hot spots at lower energies within the present two-loop RG theory, 
in agreement with previous results regarding the renormalization of the quasiparticle weight $Z$
obtained in Ref. \cite{Freire2}.

From Eq. (\ref{Grenorm}), we can also calculate here the spectral function of the model near the hot spots
$A(p_0,\mathbf{p=k_F})=-(1/\pi)\text{Im} G_{R}(p_0+i0^{+},\mathbf{p=k_F})$, which yields \cite{Ferraz}

\vspace{-0.4cm}

\begin{align}
&A(p_0,\mathbf{p=k_F})=\theta(-p_0)\left(\frac{|p_0|}{\Lambda}\right)^{\eta^{*}}\frac{\sin(\pi\eta^{*})}{\pi|p_0|},
\end{align}

\noindent and the tunneling density of states $N(p_0)=\int \frac{d^{2}\mathbf{p}}{(2\pi)^{2}}A(p_0,\mathbf{p})$
that becomes $N(p_0)\propto |p_{0}|^{\eta^{*}}$.
Thus, close to the nontrivial fixed point obtained here at two-loop RG order, 
there is a power-law suppression of the local single-particle density of states, which
could be associated with the pseudogap regime observed in underdoped cuprates. 
Indeed, the above theoretical predictions regarding the scaling forms 
of the spectral function near the hot spots and the local density of states of the present model 
could be ultimately verified experimentally using, e.g., angle-resolved photoemission spectroscopy (ARPES) 
and scanning tunneling microscope (STM) techniques. 

\section{Linear Response Theory}

To investigate what are the enhanced correlations in the model at lower energies near the hot spots, it is important to calculate 
also the flow equations for several linear response functions 
(for more details, see also Refs. \cite{Freire,Freire3} in the context
of different 2D fermionic systems). Therefore, we
must add to the Lagrangian of the model [Eq. (\ref{lagrangian})] the following term

\begin{figure}[t]
 \includegraphics[height=5.5cm]{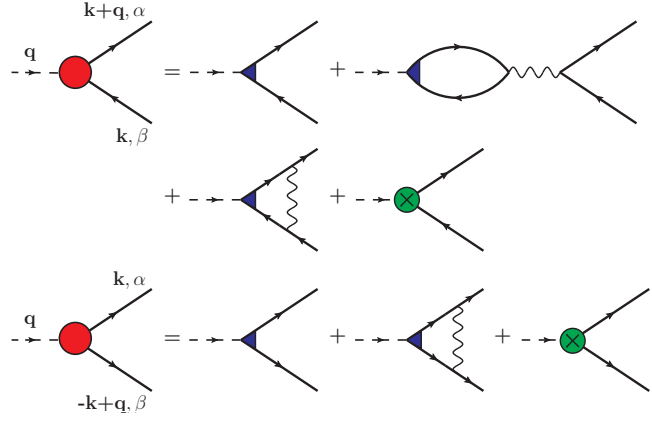}
 \caption{(Color online) Feynman diagrams for the DW and SC response
functions (three-legged vertices) of the present model which yield logarithmic divergences
as a function of the external energies. Solid lines denote noninteracting fermionic single-particle Green's function, while the wavy lines correspond to the renormalized coupling constants. 
For the DW case, the vector $\mathbf{q}$ could be either the commensurate wavevector $\mathbf{Q}=(\pi,\pi)$ 
or the incommensurate wavevector $\mathbf{\widetilde{Q}}=(Q_0,Q_0)$, depending on the channel analyzed. 
The diagrams with crosses represent the counterterms given 
by $\delta \Delta_{R,DW}^{\alpha\beta}(\mathbf{k,q})$ and $\delta \Delta_{R,SC}^{\alpha\beta}(\mathbf{k,q})$.}
 \label{fig:response_function}
\end{figure}

\vspace{-0.3cm}

\begin{align}\label{external}
&L_{ext}=\sum_{\mathbf{k},\alpha,\beta}\Delta_{B,SC}^{\alpha\beta}(\mathbf{k},\mathbf{q})\,\overline{\psi}_{B\alpha}(\mathbf{k})\overline{\psi}_{B\beta}(-\mathbf{k}+\mathbf{q}) \nonumber\\
&+\sum_{\mathbf{k},\alpha,\beta}\Delta_{B,DW}^{\alpha\beta}(\mathbf{k},\mathbf{q})\,\overline{\psi}_{B\alpha}(\mathbf{k}+\mathbf{q})\psi_{B\beta}(\mathbf{k})
+ H.c.,
\end{align}

\noindent where $\Delta_{B,SC}^{\alpha\beta}(\mathbf{k},\mathbf{q})$ and $\Delta_{B,DW}^{\alpha\beta}(\mathbf{k},\mathbf{q})$ are the bare response vertices
for the superconducting (SC) and density-wave (DW) orders, respectively. This
added term will generate new Feynman diagrams with three-legged vertices that are displayed in Fig. \ref{fig:response_function},
which will also produce new logarithmic divergences as a function of the RG scale $\Lambda$ in the model.
We must then define the renormalized
response vertices together with their corresponding counterterms
as follows: 
$\Delta_{B,i}^{\alpha\beta}(\mathbf{k},\mathbf{q})=Z^{-1}[\Delta_{R,i}^{\alpha\beta}(\mathbf{k},\mathbf{q})+\delta \Delta_{R,i}^{\alpha\beta}(\mathbf{k},\mathbf{q})]$ for $i=SC$ and $DW$. 
Then, by invoking the RG condition for bare quantities of the model $\Lambda (d\Delta^{\alpha\beta}_{B,SC(DW)}(\mathbf{k},\mathbf{q})/d\Lambda) =0$,
and either symmetrizing or antisymmetrizing these response vertices to obtain the appropriate order parameters (see Appendix C for the technical details), 
we find that the RG flow equations up to two-loop order for some physically important renormalized response vertices in the model become

\vspace{-0.3cm}

\begin{align}
\Lambda \frac{d}{d\Lambda}\Delta_{SDW}^{s-wave}&=-\frac{1}{2}(g_{2R}+g_{2x,R}+g_{3R}+g_{3x,R}+2g_{3v,R})\nonumber\\
&\times\Delta_{SDW}^{s-wave}+\eta\,\Delta_{SDW}^{s-wave},\label{eq_Susc1}\\
\Lambda\frac{d}{d\Lambda}\widetilde{\Delta}_{SDW}^{s-wave}&=-\frac{1}{2}(g_{2c,R}+g_{3p,R})\,\widetilde{\Delta}_{SDW}^{s-wave}\nonumber\\
&+\eta\,\widetilde{\Delta}_{SDW}^{s-wave},\label{eq_Susc2}\\
\Lambda\frac{d}{d\Lambda}\widetilde{\Delta}_{CDW}^{d-wave}&=\frac{1}{2}(2g_{1c,R}-g_{2c,R}+g_{3p,R}-2g_{3x,R})\nonumber\\
&\times\widetilde{\Delta}_{CDW}^{d-wave}+\eta\,\widetilde{\Delta}_{CDW}^{d-wave},\label{eq_Susc3}\\
\Lambda \frac{d}{d\Lambda}\Delta_{SSC}^{d-wave}&=\frac{1}{2}(g_{2c,R}+g_{1c,R}-g_{1x,R}-g_{2x,R}-2g_{1s,R}\nonumber\\
&+2g_{1r})\Delta_{SSC}^{d-wave}+\eta\,\Delta_{SSC}^{d-wave},\label{eq_Susc4}
\end{align}

\noindent where the tilde in some response functions is just to remind the reader that these DW vertices are calculated 
at the incommensurate wavevector $\mathbf{\widetilde{Q}}=(Q_0,Q_0)$, while the SDW response vertex without the tilde
is calculated at $\mathbf{{Q}}=(\pi,\pi)$.
Here the labels SDW($s$-wave), CDW($d$-wave) and SSC($d$-wave) refer to, respectively, spin density wave order of $s$-wave type, 
charge density wave order of $d$-wave type, and $d$-wave singlet superconductivity. 

We digress for a moment and analyze the above response functions first in the one-loop RG approximation close to the
critical scale $\Lambda_c$, which was discussed earlier in this paper. They can be calculated analytically using the scaling 
ansatz defined in Eq. \eqref{ansatz}. Substituting this expression into Eqs. \eqref{eq_Susc1}--\eqref{eq_Susc4} and 
setting the two-loop contribution from the anomalous dimension to zero (i.e. $\eta=0$), we find that the various response vertices behave near the energy scale $\Lambda_c$ as a 
power-law given by $\Delta_{m}(\Lambda)\sim(\Lambda-\Lambda_c)^{\alpha_m}$, with the exponents given approximately by

\vspace{-0.3cm}

\begin{align}
 \alpha^{s-wave}_{SDW}&=-\frac{1}{2}(C_{2}+C_{2x}+C_{3}+C_{3x}+2C_{3v})\nonumber\\
 &\approx -0.749,\\
 \widetilde{\alpha}^{s-wave}_{SDW}&=-\frac{1}{2}(C_{2c}+C_{3p})\approx -0.044,\\
 \widetilde{\alpha}^{d-wave}_{CDW}&=\frac{1}{2}(2C_{1c}-C_{2c}+C_{3p}-2C_{3x})\nonumber\\
 &\approx -0.160,\\
 \alpha^{d-wave}_{SSC}&=\frac{1}{2}(C_{2c}+C_{1c}-C_{1x}-C_{2x}-2C_{1s}\nonumber\\
 &+2C_{1r})\approx -0.186.
\end{align}

\noindent We observe from the above one-loop RG result that all exponents of the response functions analyzed in this work turn out to be negative.
Therefore, these response vertices diverge close to the
critical scale $\Lambda_c$ at this order, with the SDW($s$-wave) at $\mathbf{{Q}}=(\pi,\pi)$ (i.e. antiferromagnetic
spin correlations) being the leading ordering tendency, the $d$-wave singlet superconductivity being the subleading ordering tendency and
$d$-wave charge density wave at the incommensurate wavevector $\mathbf{\widetilde{Q}}=(Q_0,Q_0)$ coming in third place.
We draw attention to the fact that the numerical values of the exponents $\alpha^{d-wave}_{SSC}$ and $\widetilde{\alpha}^{d-wave}_{CDW}$
are also in close proximity to each order.
We argue that those findings are not accidental and they will play an important role in the low-energy effective description of the model
(to be discussed in the two-loop RG calculation below). 
It turns out to be a precursor of an emergent approximate
$SU(2)$ pseudospin symmetry in the present model. 
This emergent symmetry has been first proposed very recently in the literature in the context of the spin-fermion model
in Ref. \cite{Sachdev} and its physical consequences were explored in Refs. \cite{Efetov,Sachdev6,Sachdev2}.

The leading response function divergence at the energy scale $\Lambda_c$ in one-loop order would in principle 
indicate a phase transition at a finite temperature $T_c$ 
towards antiferromagnetism in the model, which of course contradicts the Mermin-Wagner theorem \cite{Mermin}
that states that there can be no spontaneous breaking of a continuous symmetry 
at finite temperatures in a 2D model with short-range interactions.
The above result can in fact be interpreted as an artifact of the one-loop RG approximation. 
This problem is successfully corrected by the two-loop RG approach as we will see next. For this reason, we now turn our attention to this case.

At two-loop RG order, we obtained earlier in this paper that the renormalized couplings do not display any divergence
as a function of $\Lambda$ and approach asymptotically a nontrivial 
fixed point [Eq. (\ref{fixed_point_two_loop})] in the infrared regime, which controls the universal physics of the model at large time scales and long distances. 
Indeed, by solving Eqs. \eqref{eq_Susc1}--\eqref{eq_Susc4} close to this fixed point, we obtain
that the calculated response vertices must necessarily satisfy power-laws described by $\Delta_{m}(\Lambda)\sim\Lambda^{\nu_{m}^{*}}$,
where the two-loop critical exponents are now given by

\vspace{-0.3cm}

\begin{align}
 \nu^{s-wave}_{SDW}&=-\frac{1}{2}(g^{*}_{2}+g^{*}_{2x}+g^{*}_{3}+g^{*}_{3x}+2g^{*}_{3v})\nonumber\\
 &+\eta^{*}\approx 0.041,\label{nu1}\\
 \widetilde{\nu}^{s-wave}_{SDW}&=-\frac{1}{2}(g^{*}_{2c}+g^{*}_{3p})+\eta^{*}\approx 4.222,\label{nu2}\\
 \widetilde{\nu}^{d-wave}_{CDW}&=\frac{1}{2}(2g^{*}_{1c}-g^{*}_{2c}+g^{*}_{3p}-2g^{*}_{3x})\nonumber\\
 &+\eta^{*}\approx 0.304,\label{nu3}\\
 \nu^{d-wave}_{SSC}&=\frac{1}{2}(g^{*}_{2c}+g^{*}_{1c}-g^{*}_{1x}-g^{*}_{2x}-2g^{*}_{1s}\nonumber\\
 &+2g^{*}_{1r})+\eta^{*}\approx 0.304.\label{nu4}
\end{align}

\noindent (Again, the tilde in some critical exponents is just to remind the reader that those DW vertices are calculated 
at the incommensurate wavevector $\mathbf{\widetilde{Q}}=(Q_0,Q_0)$.) We therefore
conclude from the above result that at two-loop RG level the critical exponents of all response functions calculated in this work indeed become
positive. As a consequence, instead of a divergence displayed by those response vertices at one-loop order,
they now clearly scale down to zero in the infrared regime at two loops.
This implies that there should be no spontaneous symmetry breaking at finite temperatures and
those ordering tendencies should manifest themselves at most as short-range orders
in the present system. This is consistent with the Mermin-Wagner theorem \cite{Mermin}.

From Eqs. \eqref{nu1}--\eqref{nu4}, we also note that the
antiferromagnetic fluctuations continue to be the dominant correlations in the system close to the fixed point at two loops, 
with a finite but large spin correlation length. 
Another interesting result is related to the fact that the critical exponents $\widetilde{\nu}^{d-wave}_{CDW}$ and $\nu^{d-wave}_{SSC}$ associated
with subleading orders in the system are now
approximately equal to each other, at least within the present numerical precision. As was anticipated in the previous one-loop calculation,
this is due to an emergent approximate $SU(2)$ pseudospin symmetry \cite{Sachdev} associated with particle-hole transformations 
in the model (one for each pair of hot spots
connected by $\mathbf{{Q}}=(\pi,\pi)$). This symmetry maps the $d$-wave superconducting response vertex $\Delta_{SSC}^{d-wave}$ 
onto the $d$-wave incommensurate charge order $\widetilde{\Delta}_{CDW}^{d-wave}$. Therefore, we have 
demonstrated numerically that this symmetry is indeed almost exact in the present model, insofar as the system is 
sufficiently close to the nontrivial fixed point obtained here at two-loop RG order.

To support this physical picture, we can now calculate the renormalized susceptibilities, which 
are given by

\vspace{-0.5cm}

\begin{align}\label{susc}
\chi_{m}(\Lambda)&=D(0)\int_{0}^{l} d\xi\Delta_{m}(\xi)\Delta^{*}_{m}(\xi),\\
\widetilde{\chi}_{n}(\Lambda)&=D(0)\int_{0}^{l} d\xi\widetilde{\Delta}_{n}(\xi)\widetilde{\Delta}^{*}_{n}(\xi),
\end{align}

\begin{figure}[t]
 \includegraphics[height=2.4in]{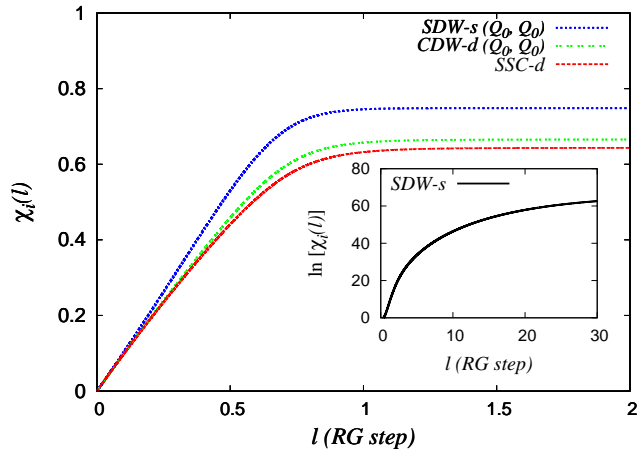}
 \caption{(Color online) Two-loop RG flow for some important susceptibilities 
 of the model (in units of $D(0)$) for the initial choice of $g_{iR}{(0)}=0.5$.}
 \label{fig:RG_susceptibilities}
\end{figure}

\noindent where $m=$ SDW($s$-wave), SSC($d$-wave) and $n=$ SDW($s$-wave), CDW($d$-wave).
The numerical results for the renormalized susceptibilities are plotted in Fig. \ref{fig:RG_susceptibilities}. 
In this figure, we confirm that even though those susceptibilities are initially enhanced at high energies, they
become saturated at plateaus for lower energies at two loops, indicating clearly short-range order. Another
important result we wish to emphasize here is that although the dominant antiferromagnetic correlations in the system are short-ranged, 
they turn out to be nearly critical close to the nontrivial fixed point obtained at two-loop RG order 
(see the logarithmic scale in the inset of Fig. \ref{fig:RG_susceptibilities}). This implies
that the emergent $SU(2)$ pseudospin symmetry that relates the $d$-wave superconducting order 
to a $d$-wave incommensurate charge order indeed takes place at lower energies in the present model 
under the crucial influence of strong short-range SDW fluctuations.

\section{Conclusions}

We have performed a RG calculation of a 2D fermionic model, which includes only the so-called hot spots
that are directly connected by SDW ordering wave vector on an underlying Fermi surface 
generated by the 2D $t-t'$ Hubbard model at low hole doping. 
By following the field-theoretical RG strategy, we have computed the
Callan-Symanzik RG equation up to two loops describing the flow of
the single-particle Green's function, the renormalized couplings, the Fermi velocity, and some of the most important order-parameter
susceptibilities in the model at lower energies. Despite the fact that at one-loop order the renormalized couplings
diverge at a finite critical scale $\Lambda_c$, the same couplings at the two-loop case 
flow to an infrared-stable nontrivial fixed point, which 
controls the universal physics of the model at large time scales and long distances. We have explored
here the implications of such a fixed point in the model. As a result,
we have obtained that the analytical structure of the renormalized Green's function of the model does not
exhibit a quasiparticle form, thus signaling a breakdown of Fermi liquid
behavior near the hot spots. We have also predicted theoretically scaling forms for the spectral function
of the model and the local density of states close to this fixed point,
which could be verified by means of angle-resolved photoemission spectroscopy and
scanning tunneling microscope experiments.

By analyzing the response functions of the model, we have established
that in the presence of strong short-range antiferromagnetic correlations in the model,
an approximate $SU(2)$ pseudospin symmetry emerges sufficiently close to the nontrivial fixed point, which relates
the $d$-wave Cooper pairing $\Delta_{SSC}^{d-wave}$ 
to the $d$-wave incommensurate charge order response $\widetilde{\Delta}_{CDW}^{d-wave}$.
This incommensurate CDW($d$-wave) order leads
to modulations in the nearest neighbor bond variables 
$<\psi^{\dagger}_{\mathbf{r\sigma}}\psi_{\mathbf{r+\hat{x},\sigma}}>$ and $<\psi^{\dagger}_{\mathbf{r}\sigma}\psi_{\mathbf{r+\hat{y}},\sigma}>$ 
but not in the charge density \cite{Sachdev}
($\hat{x}$ and $\hat{y}$ are unit vectors corresponding to the sides of
the square lattice unit cell).
For this reason, it has the character of a valence bond solid \cite{Sachdev,Sachdev4} 
(or a quadrupole-density-wave in the terminology of Ref. \cite{Efetov}). 
This conclusion shares some aspects in common with other very recent works that
obtain the same type of electronic order also emerging in 
the spin-fermion model with dominant antiferromagnetic interactions within both Hartree-Fock mean-field calculations \cite{Sachdev2} 
and also from RG methods \cite{Sachdev,Efetov}.
We point out that all of our present results
are consistent with the fact that the quasiparticle weight $Z$ is nullified near the hot spots 
and that both uniform spin and charge
susceptibilities are suppressed in the low-energy limit, which was previously obtained in Ref. \cite{Freire2}.
This may indicate either a partial truncation of the Fermi surface at the hot spots (e.g., Fermi arcs) 
or it could be also a precursor of a full reconstruction of the Fermi surface into pockets. 
Thus, the properties of the critical theory calculated here have some similarities 
with the phenomenology exhibited by the underdoped cuprates
at high temperatures \cite{PALee}.
An important challenge that we leave for a future investigation is to explore the consequences of the nontrivial fixed point 
obtained in the present work to describe the evolution of the Fermi surface 
as a function of doping observed in both the pseudogap phase and
the `strange' metal for underdoped and optimally doped cuprates, respectively.

\begin{acknowledgments}
One of us (HF) wants to thank S. Sachdev, S. Whitsitt, C. Pépin, A. Ferraz, and P. A. Lee for useful discussions.
We acknowledge financial support from CNPq under grant No. 245919/2012-0 and FAPEG under grant No. 201200550050248 for this project.
\end{acknowledgments}

\appendix

\section{}

In this appendix, we start by showing explicitly the $\beta_i$ functions that appear in the Callan-Symanzik equation (Eq. (\ref{CZE})) 
for the 2D fermionic model with eight hot spots. For simplicity, we assume that $v_{\perp R}\gg v_{\parallel R}$. 
As a result, these functions that are conventionally defined by 
$\beta_{i}=\Lambda d g_{i R}/d\Lambda$ 
become up to two-loop order

\vspace{-0.3cm}

\begin{widetext}

\begin{eqnarray}
\beta_{1}&=&g^{2}_{1}+g^{2}_{1x}+4g^{2}_{3t}+g^{2}_{3p}-g_{1x}g_{2x}-g_{3p}g_{3x}-4g_{3v}g_{3t}+\frac{1}{2}(g_{1x}g_{2x}-g^{2}_{2x}-g_{3p}g_{3x})g_{1c}\nonumber\\
&+&\frac{1}{2}(g^{2}_{1c}+g^{2}_{1}+g^{2}_{1x}+g^{2}_{2x}-g_{1x}g_{2x}-g_{3p}g_{3x}+g^{2}_{3p}+g^{2}_{3x})g_{1},\\
\beta_{2}&=&\frac{1}{2}\left(g^{2}_{1}-g^{2}_{2x}-g^{2}_{3}-g^{2}_{3x}\right)-2g^{2}_{3v}+\frac{1}{4}\left(g^{3}_{1}+g_{1c}g^{2}_{1x}+g_{1}g^{2}_{1c}\right)+\frac{1}{4}\left(2g_{2}-g_{1}\right)g^{2}_{3}\nonumber\\
&+&\frac{1}{4}\left[\left(2g_{2c}-g_{1c}\right)\left(g^{2}_{3p}+g^{2}_{3x}\right)-2g_{2c}g_{3p}g_{3x}+2g_{2}\left(g^{2}_{3p}+g^{2}_{3x}-g_{3p}g_{3x}\right)\right]\nonumber\\
&+&\frac{1}{2}(g_{2}-g_{2c})(g_{1x}^{2}+g_{2x}^{2}-g_{1x}g_{2 x}),\\
\beta_{3}&=&(g_{1}-2g_{2})g_{3}-g_{1x}(g_{3x}-2g_{3p})-g_{2x}(g_{3p}+g_{3x})+4(g_{3t}-g_{3v})g_{3t}-2g^{2}_{3v}+\frac{1}{4}[(g_{1}-2g_{2})^{2}\nonumber\\
&+&(g_{1c}-2g_{2c})^{2}+2g^{2}_{1x}+2g^{2}_{2x}-2g_{1x}g_{2x}-2g_{3p}g_{3x}+2g^{2}_{3p}+2g^{2}_{3x}+g^{2}_{3}]g_{3},\\
\beta_{3t}&=&(2g_{1}-g_{2}+g_{3}+2g_{1x}-g_{2x}+2g_{3p}-g_{3x})g_{3t}
-(g_{1}+g_{3}+g_{1x}+g_{3p})g_{3v}+2\eta g_{3t},\\
\beta_{3v}&=&-(g_{2}+g_{3}+g_{2x}+g_{3x})g_{3v}+2\eta g_{3v},
\end{eqnarray}

\begin{eqnarray}
\beta_{1c}&=&g^{2}_{1c}+g_{1x}g_{2x}+g^{2}_{1s}+g^{2}_{1r}+g^{2}_{3x}-g_{3p}g_{3x}+\frac{1}{2}\left(g_{1x}g_{2x}-g^{2}_{2x}-g_{3p}g_{3x}\right)g_{1}\nonumber\\
&+&\frac{1}{2}\left(g^{2}_{1c}+g^{2}_{1}+g^{2}_{1x}+g^{2}_{2x}-g_{1x}g_{2x} +g^{2}_{3p}+g^{2}_{3x}-g_{3p}g_{3x}\right)g_{1c},\\
\beta_{2c}&=&\frac{1}{2}\left(g^{2}_{1c}+g^{2}_{1x}+g^{2}_{2x}+2g^{2}_{1s}+2g^{2}_{1r}-g^{2}_{3p}\right)+\frac{1}{4}\left(g_{1}g^{2}_{1x}+g_{1c}g^{2}_{1}+g^{3}_{1c}\right)\nonumber\\
&+&\frac{1}{2}(g_{2c}-g_{2})(g_{1x}^{2}+g^{2}_{2x}-g_{1x}g_{2x})+\frac{1}{4}\left(2g_{2c}-g_{1c}\right)g^{2}_{3}\nonumber\\
&+&\frac{1}{4}\left[\left(2g_{2}-g_{1}\right)\left(g^{2}_{3p}+g^{2}_{3x}\right)-2g_{2}g_{3p}g_{3x}+2g_{2c}\left(g^{2}_{3p}+g^{2}_{3x}-g_{3p}g_{3x}\right)\right],\\
\beta_{1x}&=&g_{1c}g_{2x}+g_{2c}g_{1x}+2g_{1s}g_{1r}+2g_{1x}g_{1}-g_{2x}g_{1}-g_{1x}g_{2}+(g_{3p}-g_{3x})g_{3}+ 4g^{2}_{3t}-4g_{3v}g_{3t}\nonumber\\
&+&\frac{1}{2}\biggl(g_{1}g_{2c}+g_{1c}g_{2}-2g_{2c}g_{2}-\frac{g^{2}_{3p}}{2}-\frac{g^{2}_{3x}}{2}\biggr)g_{1x}+2\eta g_{1x},\\
\beta_{2x}&=&g_{1c}g_{1x}+g_{2c}g_{2x}+2g_{1s}g_{1r}-g_{2}g_{2x}-g_{3}g_{3x}-g^{2}_{3v}+\frac{1}{2}\biggl(g_{1c}g_{1}g_{1x}-2g_{1c}g_{1}g_{2x}+g_{1c}g_{2}g_{2x}\nonumber\\
&+&g_{1}g_{2c}g_{2x}-2g_{2c}g_{2}g_{2x}-g_{1x}g_{3p}g_{3x}+g_{2x}g_{3p}g_{3x}-\frac{1}{2}g_{2x}(g_{3x}^{2}+g_{3p}^{2})\biggr)+2\eta g_{2x},\\
\beta_{1s}&=&(g_{1c}+g_{2c})g_{1s}+(g_{1x}+g_{2x})g_{1r}+2\eta g_{1s},\\
\beta_{1r}&=&(g_{1c}+g_{2c})g_{1r}+(g_{1x}+g_{2x})g_{1s}+2\eta g_{1r},\\
\beta_{3p}&=&(2g_{1}-g_{2c})g_{3p}+g_{1x}g_{3}+4g^{2}_{3t}-g_{2}g_{3p}-g_{2x}g_{3}-g_{1}g_{3x}        -4g_{3v}g_{3t}+\frac{1}{2}\biggl(2g_{2c}g_{2}g_{3p}\nonumber\\
&+&g^{2}_{2x}g_{3x}-g_{1}g_{2c}g_{3p}-g_{1c}g_{2}g_{3p}-g_{1x}g_{2x}g_{3x}-g_{1c}g_{1}g_{3x}-\frac{g^{2}_{1x}g_{3p}}{2}\biggr)+2\eta g_{3p},\\
\beta_{3x}&=&(2g_{1c}-g_{2c})g_{3x}-g_{1c}g_{3p}-g_{2x}g_{3}-g_{2}g_{3x}-g^{2}_{3v}+\frac{1}{2}\biggl(2g_{2c}g_{2}g_{3x}+g^{2}_{2x}g_{3p}-g_{1}g_{2c}g_{3x}\nonumber\\
&-&g_{1c}g_{2}g_{3x}-g_{1x}g_{2x}g_{3p}-g_{1c}g_{1}g_{3p}-\frac{g^{2}_{1x}g_{3x}}{2}\biggr)+2\eta g_{3x},
\end{eqnarray}

\end{widetext}

\noindent  where we have suppressed the indices $R$ in the renormalized dimensionless couplings not to clutter up the notation
and $\eta$ is the anomalous dimension contribution. 
Those RG beta functions $\beta_i$  
describe the flow of the effective couplings of this model in a thirteen-dimensional coupling parameter space 
as the RG scale $\Lambda$ is lowered continuously. The corresponding Feynman diagrams up to two-loop order
are depicted below in Fig. \ref{fig:Vertex_corrections}. 

\begin{figure}[t]
 \includegraphics[height=5.8cm]{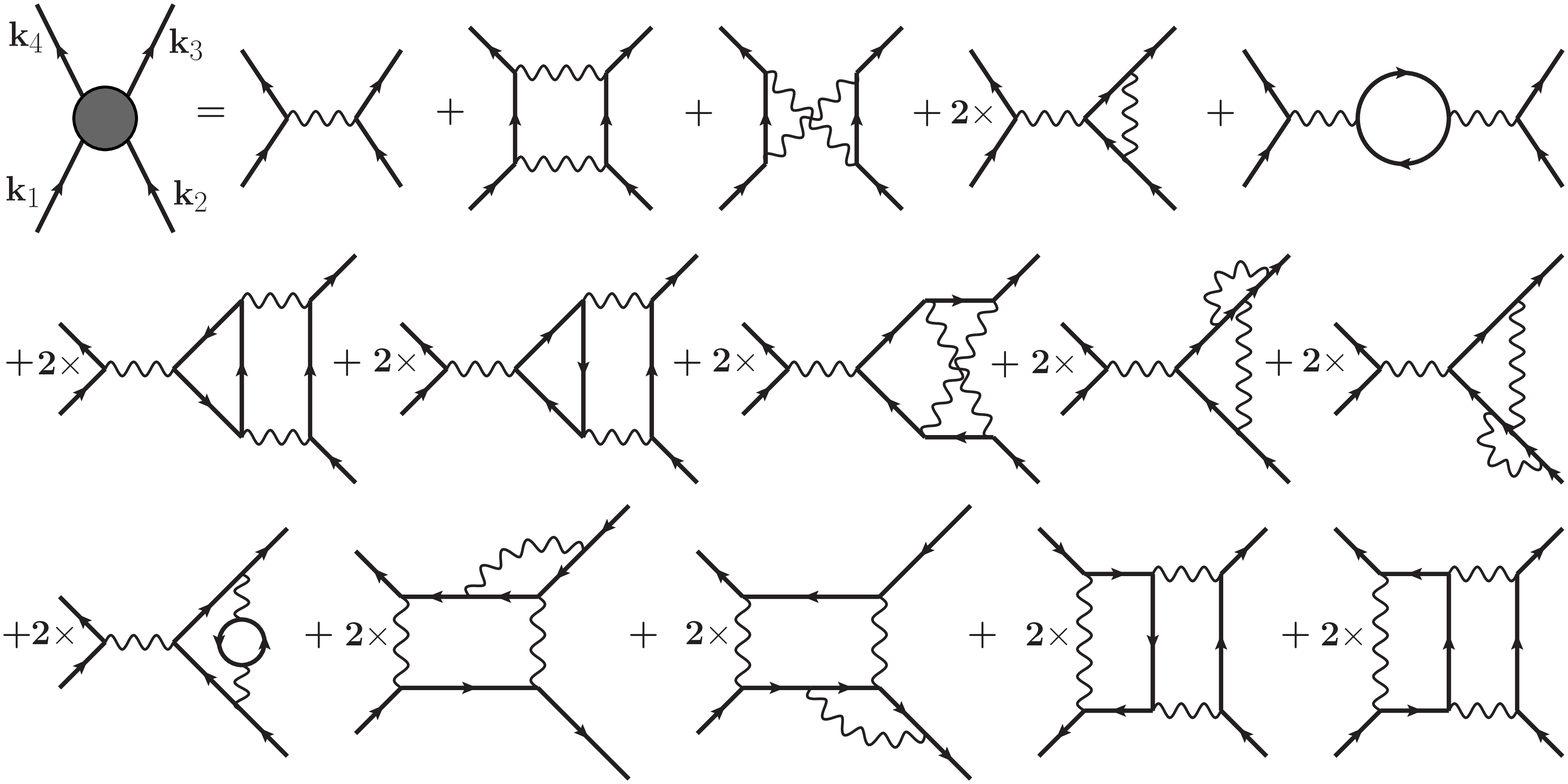}\\
 \includegraphics[height=2cm]{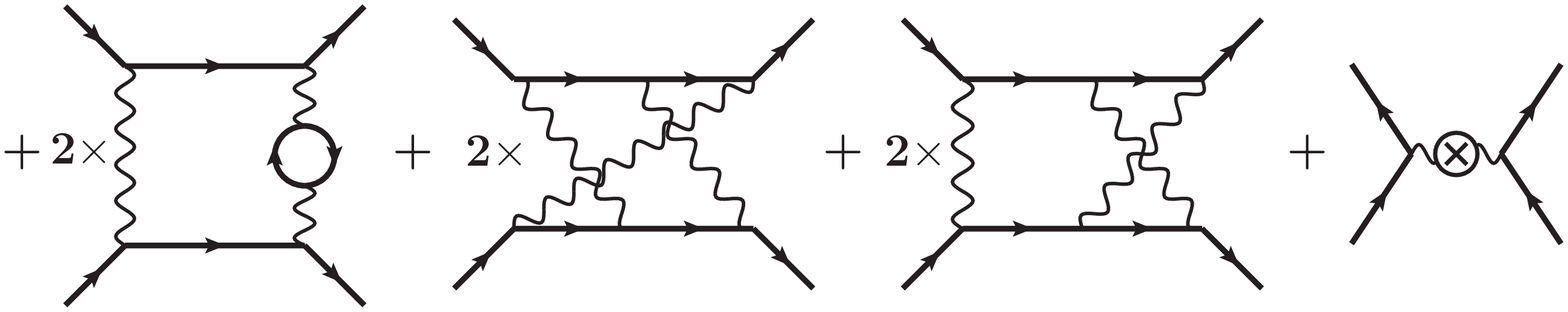}
 \caption{Feynman diagrams for the vertex corrections up
to two-loop order which yield logarithmic divergences as a
function of the external energies. Solid lines denote noninteracting fermionic single-particle Green's function, 
while the wavy lines correspond to the renormalized coupling constants. 
The diagram with a cross refers to the counterterm (given by $\delta g_{i R}$) in
each corresponding scattering channel.}
 \label{fig:Vertex_corrections}
\end{figure}

\section{}

Next, we move on to calculate the anomalous dimension defined by $\eta=\Lambda \,d\ln Z/d\Lambda$
that describes the renormalization
of the quasiparticle weight $Z$ as a function of the RG scale $\Lambda$, and
the functions $\eta_{v_{\perp}}=\Lambda\, d\ln Z_{v_{\perp}}/d\Lambda$ and 
$\eta_{v_{\parallel}}=\Lambda\, d\ln Z_{v_{\parallel}}/d\Lambda$ which are related to the RG flows of the 
components of Fermi velocity given by $\mathbf{v_F}$=($v_{\parallel}$,$v_{\perp}$). For the $Z$ factor,
using the standard RG condition \cite{Freire2} for
the inverse of renormalized single-particle Green's function $\Gamma^{(2)}_{R}\equiv(G_{R})^{-1}=(G_{0})^{-1}-\Sigma_{R}$, namely
$\text{Re}\,\Gamma^{(2)}_{R}(p_0=\Lambda,|p_{\parallel}|=k_{F}^{\parallel},|p_{\perp}|=k_{F}^{\perp})=\Lambda$, and using Eq. (\ref{self-energy}),
we obtain that $\delta Z= Z-1= (\gamma/4)\ln(\Lambda/\Lambda_0)$ from which $\eta=\gamma/4$ follows, where
$\gamma=(g^{2}_{1R}+g^{2}_{2R}+g^{2}_{1cR}+g^{2}_{2cR}+g^{2}_{1xR}+g^{2}_{2xR}
-g_{1R}g_{2R}-g_{1cR}g_{2cR}-g_{1xR}g_{2xR}-g_{3pR}g_{3xR}+g^{2}_{3pR}
+g^{2}_{3xR}+\frac{g^{2}_{3R}}{2})$.
As for the component $v_{\perp}$, it can be advantageous to use the RG prescription, e.g., for half of the hot spots, as follows
$\text{Re}\,\Gamma^{(2)}_{R}(p_0=0,|p_{\parallel}|=k_{F}^{\parallel},v_{\perp R}(|p_{\perp}|-k_{F}^{\perp})=\Lambda)=-\Lambda$
(we point out that, for the other half of the hot spots, the role of the components $p_{\parallel}$ and $p_{\perp}$ is of course interchanged). Therefore, 
using Eq. (\ref{self-energy}), we obtain $\delta Z_{v_{\perp}}= Z_{v_{\perp}}-1= (\gamma/4)\ln(\Lambda/\Lambda_0)$ and $\eta_{v_{\perp}}=\gamma/4$. Since
$v_{\perp B}=Z_{v_{\perp}}Z^{-1}v_{\perp R}$, it follows that the $v_{\perp R}$ is independent of the RG scale $\Lambda$, i.e.
$\Lambda\, (d v_{\perp R}/d\Lambda)=0$ and,
as a result, the normal component of the Fermi velocity at any hot spot is always an RG invariant at this order of perturbation theory. 
Lastly, in an analogous way for the component $v_{\parallel}$,
it can be convenient to choose the RG condition, e.g., for half of the hot spots, as follows
$\text{Re}\,\Gamma^{(2)}_{R}(p_0=0,v_{\parallel R}(|p_{\parallel}|-k_{F}^{\parallel})=\Lambda,
|p_{\perp}|=k_{F}^{\perp})=\Lambda$,
from which $\eta_{v_{\parallel}}=\gamma/4$ follows straightforwardly (we mention that the same remark regarding
the role of $p_{\parallel}$ and $p_{\perp}$ depending on the hot spot analyzed also applies here). 
In view of the fact that $v_{\parallel B}=Z_{v_{\parallel}}Z^{-1}v_{\parallel R}$,
this means that $v_{\parallel R}$ is also not renormalized at this order of perturbation theory. 

\section{}

Now we derive the RG flow equations up to two-loop order for the renormalized response vertices.  
By calculating the Feynman diagrams displayed in Fig. \ref{fig:response_function} (assuming, for simplicity, that $v_{\perp R}\gg v_{\parallel R}$) and 
invoking the RG condition for the bare quantities of the model $\Lambda (d\Delta^{\alpha\beta}_{B,SC(DW)}(\mathbf{k,q})/d\Lambda) =0$,
we obtain the RG flow equations for the response vertices. We begin with the SC response vertices, which are given by

\vspace{-0.2cm}

\begin{align}\label{SC} 
&\Lambda\frac{d\Delta_{R,SC}^{(1)\alpha\beta}}{d\Lambda}=\frac{1}{2}\left[(g_{2c}+g_{1r})\Delta_{R,SC}^{(1)\alpha\beta}
-(g_{1c}+g_{1r})\Delta_{R,SC}^{(1)\beta\alpha}\right.\nonumber\\
&\left.+(g_{2x}+g_{1s})\Delta_{R,SC}^{(2)\alpha\beta}-(g_{1x}+g_{1s})\Delta_{R,SC}^{(2)\beta\alpha}\right]+\eta\,\Delta_{R,SC}^{(1)\alpha\beta},\nonumber\\
\end{align}

\noindent where we defined the new vertices $\Delta_{R,SC}^{(1)\alpha\beta}=\Delta_{R,SC}^{\alpha\beta}(k_{F}^{\parallel},k_{F}^{\perp};\mathbf{q}=0$)
and $\Delta_{R,SC}^{(2)\alpha\beta}=\Delta_{R,SC}^{\alpha\beta}(-k_{F}^{\parallel},k_{F}^{\perp};\mathbf{q}=0)$ depending on their location at the Fermi points. 
As for the DW response vertices at the commensurate wavevector $\mathbf{Q}=(\pi,\pi)$, we get the following RG equations

\vspace{-0.2cm}

\begin{align}\label{DW}
&\Lambda\frac{d\Delta_{R,DW}^{(1)\alpha\beta}}{d\Lambda}=\frac{1}{2}\bigg[(g_1+g_3+2g_{3t})\sum_{\sigma=\alpha,\beta}\Delta_{R,DW}^{(1)\sigma\sigma}\nonumber\\
&+(g_{1x}+g_{3p}+2g_{3t})\sum_{\sigma=\alpha,\beta}\Delta_{R,DW}^{(2)\sigma\sigma}-(g_{3}+g_{3x})\Delta_{R,DW}^{(1)\alpha\beta}\nonumber\\
&-(g_{2}+g_{2x})\Delta_{R,DW}^{(2)\beta\alpha}\bigg]+\eta\,\Delta_{R,DW}^{(1)\alpha\beta},
\end{align}

\noindent where we also introduced the new vertices 
$\Delta_{R,DW}^{(1)\alpha\beta}=\Delta_{R,DW}^{\alpha\beta}(k_{F}^{\parallel},k_{F}^{\perp};\mathbf{q}=\mathbf{Q})$
and $\Delta_{R,DW}^{(2)\alpha\beta}=\Delta_{R,DW}^{\alpha\beta}(-k_{F}^{\parallel},k_{F}^{\perp};\mathbf{q}=\mathbf{Q})$ according to their association to the hot spots.
Finally, for the DW response vertices at the incommensurate wavevector $\mathbf{\widetilde{Q}}=(Q_0,Q_0)$ 
(see Fig. 1) we get the following RG equations

\vspace{-0.2cm}

\begin{align}\label{DW}
&\Lambda\frac{d\widetilde{\Delta}_{R,DW}^{(1)\alpha\beta}}{d\Lambda}=\frac{1}{2}\bigg[g_{1c}\sum_{\sigma=\alpha,\beta}
\widetilde{\Delta}_{R,DW}^{(1)\sigma\sigma}
+g_{3x}\sum_{\sigma=\alpha,\beta}\widetilde{\Delta}_{R,DW}^{(2)\sigma\sigma}\nonumber\\
&-g_{2c}\widetilde{\Delta}_{R,DW}^{(1)\alpha\beta}-g_{3p}\widetilde{\Delta}_{R,DW}^{(2)\beta\alpha}\bigg]+\eta\,\widetilde{\Delta}_{R,DW}^{(1)\alpha\beta},
\end{align}

\noindent where we used the notation for the vertices $\widetilde{\Delta}_{R,DW}^{(1)\alpha\beta}=\Delta_{R,DW}^{\alpha\beta}
(k_{F}^{\parallel},k_{F}^{\perp};\mathbf{q}=\mathbf{\widetilde{Q}})$
and $\widetilde{\Delta}_{R,DW}^{(2)\alpha\beta}=\Delta_{R,DW}^{\alpha\beta}(-k_{F}^{\parallel},k_{F}^{\perp};\mathbf{q}=\mathbf{\widetilde{Q}})$
and the tilde is just to emphasize that these DW vertices are calculated at the incommensurate wavevector $\mathbf{\widetilde{Q}}=(Q_0,Q_0)$ 
(see Fig. \ref{fig:Fermi_surface}).

The last step consists of symmetrizing (or antisymmetrizing) these response vertices with respect to the spin indices. As a result, we obtain the following order parameters

\begin{center}
$\left\{%
\begin{array}{ll}
    \Delta^{(j)}_{SDW}=\Delta_{R,DW}^{(j)\uparrow\uparrow}-\Delta_{R,DW}^{(j)\downarrow\downarrow},\\ 
    \widetilde{\Delta}^{(j)}_{SDW}=\widetilde{\Delta}_{R,DW}^{(j)\uparrow\uparrow}-\widetilde{\Delta}_{R,DW}^{(j)\downarrow\downarrow},\\
    \widetilde{\Delta}^{(j)}_{CDW}=\widetilde{\Delta}_{R,DW}^{(j)\uparrow\uparrow}+\widetilde{\Delta}_{R,DW}^{(j)\downarrow\downarrow},\\
    \Delta^{(j)}_{SSC}=\Delta_{R,SC}^{(j)\uparrow\downarrow}-\Delta_{R,SC}^{(j)\downarrow\uparrow},
    \end{array}%
\right.$ 
\end{center}

\noindent where $j=1,2$ and the subscripts $SDW$ and $CDW$ stand for charge and spin density wave orders, respectively, 
whereas the subscript $SSC$ corresponds to singlet superconductivity. Therefore, using the above relations
one can derive straightforwardly the RG flow equations for each response vertex associated with a potential instability of the
normal state towards a given ordered (i.e. symmetry-broken) phase. To determine the symmetry of
the order parameter we must further symmetrize the response vertices with respect to the index $j$. Thus

\begin{center}
$\left\{%
\begin{array}{ll}
    \Delta_{SDW}^{(s-wave)}=\Delta_{SDW}^{(1)}+\Delta_{SDW}^{(2)},\vspace{0.1cm} \\
    \widetilde{\Delta}_{SDW}^{(s-wave)}=\widetilde{\Delta}_{SDW}^{(1)}+\widetilde{\Delta}_{SDW}^{(2)},\vspace{0.1cm} \\
    \widetilde{\Delta}_{CDW}^{(d-wave)}=\widetilde{\Delta}_{CDW}^{(1)}-\widetilde{\Delta}_{CDW}^{(2)},\vspace{0.1cm} \\
    \Delta^{(d-wave)}_{SSC}=\Delta_{SSC}^{(1)}-\Delta_{SSC}^{(2)},\vspace{0.1cm} \\
\end{array}%
\right.$ 
\end{center}

\noindent where the response functions stand for, respectively,
$s$-wave spin density wave ($\Delta_{SDW}^{(s-wave)}$) at the commensurate
wavevector $\mathbf{Q}=(\pi,\pi)$, $s$-wave spin density wave ($\widetilde{\Delta}_{SDW}^{(s-wave)}$) at the incommensurate
wavevector $\mathbf{\widetilde{Q}}=(Q_0,Q_0)$, $d$-wave charge density wave ($\widetilde{\Delta}_{CDW}^{(d-wave)}$) at the 
incommensurate wavevector $\mathbf{\widetilde{Q}}=(Q_0,Q_0)$ 
and $d$-wave singlet superconductivity 
($\Delta^{(d-wave)}_{SSC}$).

\end{document}